\documentclass[twocolumn,prl,showpacs,superscriptaddress,reprint]{revtex4-1}
\usepackage{graphicx}
\usepackage{epstopdf}
\usepackage{braket}
\usepackage{amsmath}
\usepackage{amssymb}
\usepackage{hyperref}
\usepackage{comment}
\pdfoutput=1

\begin{document}
	\title{Twist-controlled Resonant Tunnelling between Monolayer and Bilayer Graphene.}
	\author{T.~L.~M.~Lane}
	\email{tomlmlane@gmail.com}
	\affiliation{School of Physics \& Astronomy, University of Manchester, Oxford Road, Manchester, M13 9PL, UK}
	\author{J.~R.~Wallbank}
	\affiliation{National Graphene Institute, University of Manchester, Booth St. E, Manchester, M13 9PL, UK}
	\author{V.~I.~Fal'ko}
	\affiliation{School of Physics \& Astronomy, University of Manchester, Oxford Road, Manchester, M13 9PL, UK}
	\affiliation{National Graphene Institute, University of Manchester, Booth St. E, Manchester, M13 9PL, UK}
	
	\begin{abstract}
		\centering We investigate the current-voltage characteristics of a field-effect tunnelling transistor comprised of both monolayer and bilayer graphene with well-aligned crystallographic axes, separated by three layers of hexagonal boron nitride.  Using a self-consistent description of the device's electrostatic configuration we relate the current to three distinct tunable voltages across the system and hence produce a two-dimensional map of the I-V characteristics in the low energy regime.  We show that the use of gates either side of the heterostructure offers a fine degree of control over the device's rich array of characteristics, as does varying the twist between the graphene electrodes.
	\end{abstract}
	
	\maketitle
	
	Recently it has been demonstrated that van der Waals heterostructures of graphene and hexagonal boron nitride (hBN) \cite{lee_appPhysLett_2011, britnell_nanoLett_2012, britnell_science_2013, yang_science_2012, haigh_natMat_2012, gorbachev_natPhys_2012, ponomarenko_natPhys_2011} can be used to create tunnelling transistors \cite{georgiou_natNano_2013, ponomarenko_appPhys_2013, britnell_natComm_2013, mishchenko_nature_2014, fallahazad_nanoLet_2009, britnell_science_2012}.  Most notably, the highest quality graphene-hBN vdW structures, with ballistic electron propagation at the micron length-scale, enable one to exploit the unique crystalline structure and conductive properties of graphene \cite{geim_natMat_2007, neto_revModPhys_2009, mccann_repProgPhys_2013} in order to construct transistors featuring highly controllable I-V characteristics \cite{mishchenko_nature_2014, yang_nano_2010, wang_nanoLett_2015}.
	
	In particular, the work of Mishchenko \textit{et.~al.}~\cite{mishchenko_nature_2014} demonstrated the possibility of producing vertical-tunnelling field-effect transistors featuring a pair of graphene electrodes with well aligned crystallographic axes (misaligned by $\theta\approx1^{\circ}$).  Such devices exhibit strong resonant peaks in their current characteristics which precipitate the onset of negative differential conductance (NDC) which has been used to generate radio frequency oscillations when connected to an LC circuit \cite{gaskell_arxiv_2015}.
	
\begin{figure}[!htbp]
		\includegraphics[width=0.48\textwidth]{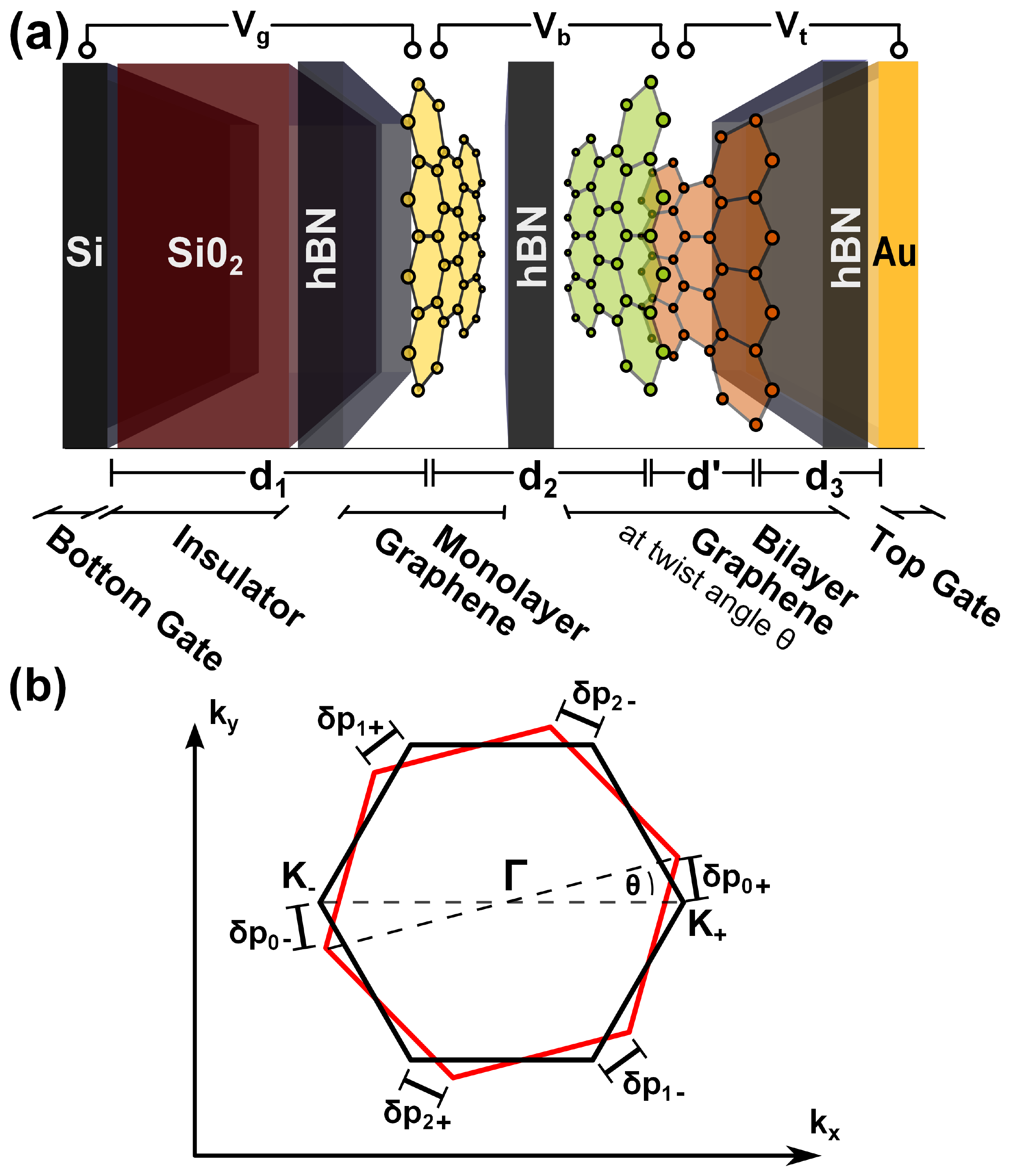}
		\caption{
			(a) Sketch of the tunnelling device indicating the voltage contacts and characteristic distances $d_{1},\ d_{2},\ d'\ \&\ d_{3}$.  Here, $V_{g},\ V_{b}$ and $V_{t}$ are the back gate, bias and top gate voltage respectively.
			(b) Momentum space map showing the orientation of the monolayer's first BZ (black) and the bilayer layer's first BZ (red) indicating the impact of the misalignment angle between them. \label{theStackWithBZ}}
	\end{figure}

	Here, we show that the resonant tunnelling characteristics of a tunnelling transistor with graphene electrodes, one monolayer and another bilayer (see sketch in Fig.\ref{theStackWithBZ}(a)), have rich I-V characteristics and display great sensitivity to the alignment of the crystal layers.  Past works have shown that we wield an exceptional amount of control over the precise electronic composition of bilayer graphene \cite{zhang_nature_2009, ohta_scienceMag_2006} due to the finely adjustable band gap within its structure.
	
	In the device modelled, graphene/hBN/bilayer-graphene is encapsulated on both top and bottom by additional multi-layers of hBN, which has been shown to increase the electronic quality of graphene layers \cite{dean_natNano_2010} and is placed on an oxidised silicon substrate which acts as a back-gate.  Further electrostatic control can be achieved with the inclusion of a top-gate.
	
	Figure \ref{theStackWithBZ}(b) shows how the real-space in-plane twist angle between the graphene flakes affects their first Brillouin zone (BZ) orientation in reciprocal space \cite{bistritzer_physRevB_2010, mele_physRevB_2011, santos_physRevB_2012,kindermann_physRevB_2012,wallbank_thesis, moon_physRevB_2014}.  A momentum shift,
	\begin{equation}
	\delta\vec{p}_{n,\xi}\approx\theta\vec{l}_{z}\times\vec{K}_{n,\xi},
	\end{equation}
	where $\xi=\pm1$ indexes two inequivalent valleys $\vec{K}_{+}/\vec{K}_{-}$ with $n=0,1,2$ indexing three equivalent K-points lying in each, is induced between the Dirac points, altering the conditions that must be met to simultaneously conserve both energy and momentum.
	
	We describe this tunnelling system using the Hamiltonian,
	\begin{subequations}
	\begin{equation}
	\hat{\mathcal H} = \hat{\mathcal H}_{ML} + \hat{\mathcal H}_{BL} + \hat{\mathcal H}_{T},
	\label{tunnelling_hamiltonian}
	\end{equation}
	where $\hat{H}_{ML}$ describes an isolated monolayer, $\hat{H}_{BL}$ describes an isolated bilayer and $\hat{H}_{T}$ characterises the interaction between the two.  For the graphene monolayer we use
	\begin{equation}
	\hat{\mathcal H}_{ML}=\sum_{\vec{k},\xi}\left(\begin{array}{c c}
	a_{\vec{k}}^{\dagger} & b_{\vec{k}}^{\dagger}
	\end{array}\right)
	\left(\begin{array}{c c}
	0 & v\hat{\pi}^{\dagger}\\
	v\hat{\pi} & 0
	\end{array}\right)
	\left(\begin{array}{c}
	a_{\vec{k}} \\ b_{\vec{k}}
	\end{array}\right).
	\label{monolayer_contribution}
	\end{equation}
	Here, $a_{\vec{k},\xi}$/$b_{\vec{k},\xi}$ are the annihilation operators for Bloch wavefunctions formed from carbon $p_z$ orbitals on the $A$/$B$ sub-lattice, $v=10^{8}cms^{-1}$ is the Dirac velocity and $\hat{\pi}=\xi k_{x} + ik_{y}$, where $\vec{k}$ is the in-plane valley momentum.  The corresponding eigenvalues are
	\begin{equation}
	\varepsilon_{ML,s_{1}} = s_{1}v|\vec k|,\label{monolayer_eigenvalue}
	\end{equation}
	\end{subequations}
	where $s_{1}=\pm1$ indexes the conduction/valence band states.  For the bilayer we use
	\begin{subequations}
	\begin{equation}
	\hat{\mathcal H}_{BL}=\sum_{\vec{k},\xi}\left(\begin{array}{c c c c}
	A_{\vec{k},\xi}^{\dagger} & B_{\vec{k},\xi}^{\dagger} & A^{'\dagger}_{\vec{k},\xi} & B^{'\dagger}_{\vec{k},\xi}
	\end{array}\right)
	\hat{\mathcal H}_{BL,\xi}^{4\times4}
	\left(\begin{array}{c}
	A_{\vec{k},\xi} \\ B_{\vec{k},\xi} \\ A_{\vec{k},\xi}^{'} \\ B_{\vec{k},\xi}^{'}
	\end{array}\right),
	\label{bilayer_contribution}
	\end{equation}
	where $A_{\vec{k},\xi}$/$B_{\vec{k},\xi}$ and $A_{\vec{k},\xi}^{'}$/$B_{\vec{k},\xi}^{'}$ are annihilation operators for Bloch wavefunctions on the $A$/$B$ carbon lattice sites for the upper and lower layer of the bilayer.  The matrix $\hat{\mathcal H}_{BL,\xi}^{4\times4}$ is given by
	\begin{equation}
	\hat{\mathcal H}_{BL,\xi}^{4\times4} = \left(\begin{array}{c c c c}
	\frac{\Delta}{2}+u & v\hat{\pi}^{\dagger} & 0 & 0 \\
	v\hat{\pi} & \frac{\Delta}{2}+u & \gamma_{1} & 0 \\
	0 & \gamma_{1} & -\frac{\Delta}{2}+u & v\hat{\pi}^{\dagger} \\
	0 & 0 & v\hat{\pi} & -\frac{\Delta}{2}+u
	\end{array}\right),
	\label{bilayer_hamiltonian}
	\end{equation}
	with $\Delta$ giving the energy difference between the two layers (the band gap), $u$ expressing the energy difference between the monolayer Dirac point and the centre of the bilayer band gap (the band offset) and $\gamma_{1}\approx0.39eV$ being the inter-layer coupling energy \cite{novoselov_natPhys_2006}.  This Hamiltonian has corresponding eigenvalues
	\begin{multline}
	\varepsilon_{BL, s_{2}, s_{3}} = u + s_{2}\biggl(s_{3}\sqrt{\Delta^{2}v^{2}|\hat{\pi}|^{2} + \frac{\gamma_{1}^{4}}{4} + \gamma_{1}^{2}v^{2}|\hat{\pi}|^{2}}\\+\frac{\Delta^{2}}{4}+v^{2}|\hat{\pi}|^{2}+\frac{\gamma_{1}^{2}}{2}\biggr)^{\frac{1}{2}},\label{bilayer_eigenvalue}
	\end{multline}
	\end{subequations}
	where $s_{2}=\pm1$ indexes the conduction/valence bands and $s_{3}=\pm1$ indexes the choice of high (split at $\pm \gamma_1$) and low (degenerate near neutrality point) energy bands.  Plotting the four resulting surfaces in momentum space produces the `Mexican hat' band structure \cite{mccann_physRevB_2006}.\\
	
	To obtain the tunnelling Hamiltonian $\hat{\mathcal{H}}_{T}$, we take the crystallographic directions of the hBN layer to be highly misaligned from the two graphene layers.  
	Thus, any tunnelling process involving scattering by hBN reciprocal lattice vectors is unable to scatter graphene's electrons between the vicinity of the BZ corners on the two layers.  This would instead result in scattering to high energy regions of graphene's BZ ( $|\varepsilon|\gg|\mu_{ML}|, |\mu_{BL}|$ ) which do not contribute to tunnelling \cite{wallbank_thesis,footnote}. Because of this we replace the hBN layer with a homogeneous insulator. We then assume that the tunnelling matrix element is controlled by the overlap between the tails of the carbon $p_{z}$ orbitals on the two layers.  Thus,
	\begin{subequations}
	\begin{multline}
	\hat{\mathcal H}_{T} = \sum_{\vec{k},\xi,n}\biggl\{
	\hat{\mathcal H}_{T,\xi}^{aA^{'}}a_{\vec{k}^{'},\xi}^{\dagger}A_{\vec{k},\xi}^{'} +
	\hat{\mathcal H}_{T,\xi}^{aB^{'}}a_{\vec{k}^{'},\xi}^{\dagger}B_{\vec{k},\xi}^{'} +\\
	\hat{\mathcal H}_{T,\xi}^{bA^{'}}b_{\vec{k}^{'},\xi}^{\dagger}A_{\vec{k},\xi}^{'} +
	\hat{\mathcal H}_{T,\xi}^{bB^{'}}b_{\vec{k}^{'},\xi}^{\dagger}B_{\vec{k},\xi}^{'}\biggr\} + h.c. ,
	\label{tunnelling_contribution}
	\end{multline}
	where $\vec{k}^{'}=\vec{k}+\delta\vec{p}_{n,\xi}$.  The matrix elements are
	\begin{eqnarray}
	\hat{H}_{T,\xi}^{\alpha\beta} &=& E_{0}\braket{\Phi_{ML,\xi}^{\alpha}(\vec{k}^{'})|\Phi_{BL,\xi}^{\beta}(\vec{k})}\nonumber\\
	&=& \Gamma e^{-i\frac{2\pi\xi n}{3}\hat{1}_{\alpha,\beta}},\label{overlap}
	\end{eqnarray}
	\end{subequations}
	where $\Phi_{ML/BL,\xi}^{\alpha/\beta}$ are the Block wavefunctions on the graphene monolayer/bilayer, $E_{0}$ is the hopping integral energy between sites, $\hat{1}_{a,A^{'}}=\hat{1}_{b,B^{'}}=0$,  $\hat{1}_{a,B^{'}}=-\hat{1}_{b,A^{'}}=1$, and we have drawn together all constant factors into $\Gamma$.  Also, note that we only include the terms which describe interaction between the monolayer and nearest (primed) layer in the bilayer.  This is because the furthest layer is separated from the monolayer by a greater distance resulting in the suppression of any tunnelling into it.
	
	After using the Fermi golden rule, we find an expression for tunnelling current density,
	\begin{widetext}
	\begin{equation} 
	I=\Gamma^{2}
	\sum_{s_{1},s_{2},s_{3},\xi,n}\int d\vec{k}\Biggl\{
	\biggl(f(\mu_{ML})-f(\mu_{BL})\biggr)
	\frac{1}{\pi}Im\left[
	\frac{\left|G_{ML}\cdot G_{BL}\right|^{2}}
	{\varepsilon_{BL,s_{2},s_{3}}-\varepsilon_{ML,s_{1}}-i\alpha}
	\right]\Biggr\},\label{fermi_golden_rule}
	\end{equation}\\[-8mm]
	\begin{equation}
	\text{where}\qquad G_{ML} = 1+\xi s_{1}e^{i\left(\xi\phi_{\vec{k}+\delta\vec{p}_{n,\xi}}-\frac{2\xi\pi n}{3}\right)}, \qquad G_{BL}= \tilde{N}\left(\frac{\Delta}{2}+\varepsilon_{BL,s_{2},s_{3}}+ v\hat{\pi}e^{-i\frac{2\xi\pi n}{3}}\right).\nonumber\label{chiral}
	\end{equation}
	\end{widetext}
	Here, $f(\mu_{ML})$ and $f(\mu_{BL})$ are the occupancy factors for the graphene monolayer and bilayer and we use $\alpha = 0.005eV$ as an energy broadening parameter within our Lorentzian giving the energy bands a finite width (perfect energy conservation is obtained in the limit $\alpha\rightarrow0$). The factors $G_{ML}$ and $G_{BL}$ arise from the sublattice composition of the graphene monolayer and bilayer wavefunctions, with $\phi_{\vec k}=\text{arctan}(k_y/k_x)$ and $\tilde{N}$ being the normalisation of the bilayer eigenvector (obtained from diagonalisation of $\hat{\mathcal{H}}_{BL}$ with the choice of $v\hat{\pi}$ amplitude for the $B^{'}$ component).  Due to the time reversal symmetry, in the absence of any external magnetic field, the valleys produce identical contributions to tunnelling current.
	
	The parameters $\mu_{ML}$, $\mu_{BL}$, $\Delta$ and $u$ in our expression for current density are calculated from a four-plate capacitor model for the two graphene layers and the gate electrodes.  Using Gauss' law we find expressions relating the static electric fields between subsequent layers to the charge on each layer.  Then, by considering a diagram of our energy bands, we may equate each of the tunable voltages' induced energy differences to our missing variables.  This produces the following four equations to be solved numerically for $\mu_{ML},\ \mu_{BL},\ \Delta$ and the charge density on the back gate, $n_{Si}$.
	\begin{subequations}
	\begin{eqnarray}
	eV_{b} &=& u - \mu_{ML} + \mu_{BL}, \label{bias_voltage_eqn}\\
	eV_{t} &=& -\frac{e^{2}d_{3}(n_{Si}+ n_{ML}(\mu_{ML})+n_{BL}(\mu_{BL}, \Delta))}{\epsilon_{0}\epsilon_{hBN}}-\mu_{BL},\nonumber\label{top_voltage_eqn}\\
	eV_{g} &=& -\frac{e^{2}d_{1}n_{Si}}{\epsilon_{0}\epsilon_{SiO_{2}}} + \mu_{ML},\nonumber\label{gate_voltage_eqn}\\
	\Delta &=& -\frac{e^{2}d_{3}(n_{Si}+ n_{ML}(\mu_{ML})+\frac{n_{BL}(\mu_{BL}, \Delta)}{2})}{\epsilon_{0}},\nonumber\label{bandGap_eqn}
	\end{eqnarray}  
	Here, $n_{ML}$ and $n_{BL}$ are the charge densities on the monolayer and bilayer obtained consistently with Eq.~(\ref{monolayer_eigenvalue}) and Eq.~(\ref{bilayer_eigenvalue}) respectively, $e$ is the electron charge, $\epsilon_{hBN}\approx\epsilon_{SiO_{2}}\approx4$ are the dielectric constants of the hBN and SiO$_{2}$ wedges and $u$ is obtained using
	\begin{equation}
	u = -\frac{e^{2}d_{2}(n_{Si}+n_{ML}(\mu_{ML}))}{\epsilon_{0}\epsilon_{hBN}}.\label{bandOffset_eqn}
	\end{equation}
	\end{subequations}
	Parameters $d_{1},\ d_{2}\ \&\ d_{3}$ define the distances between; Si substrate and monolayer, monolayer and bilayer, and bilayer and top gate, whilst $d'$ is the interlayer separation within the bilayer.  The formula for $\Delta$ assumes an equal distribution of charge between the layers of the graphene bilayer.
	
	\begin{figure*}[!hbtp]
		\includegraphics[width=\textwidth]{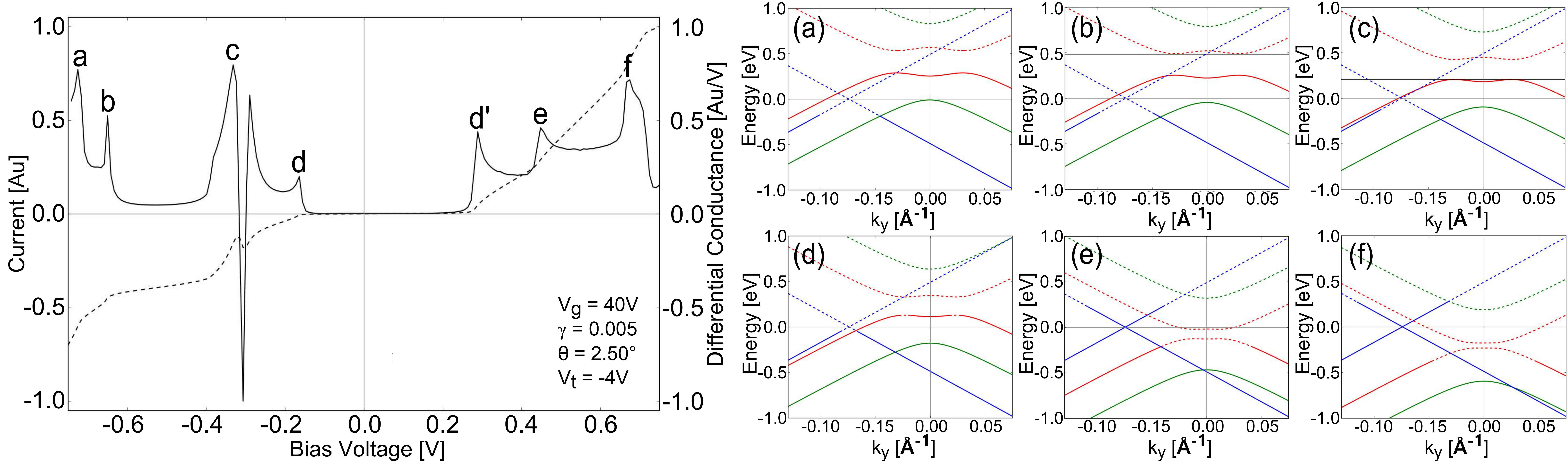}
		\caption{
			Current and its differential for $V_{g} = 40V$, $V_{t} = -4V$,	$\theta=2.5^{\circ}$, $\epsilon_{SiO_{2}}\approx\epsilon_{hBN}\approx4$,$d_{1}=3200\text{\AA}$, $d_{2}=13.5\text{\AA}$, $d_{3}=150\text{\AA}$ and $d'=3.4\text{\AA}$ (left) with all distinctive features in the differential conductance related to diagrams (right) indicating the band alignment along the 
			$k_{y}$	direction for each particular feature.  The blue lines represent the monolayer Dirac cone, the red and green curves represent the low- and high-energy bilayer bands respectively.  Solid/dashed lines distinguish between occupied/empty bands.
			\label{current_cut}}
	\end{figure*}
	
	Figure \ref{current_cut} shows the results of this model varying with $V_{b}$.  The dashed line indicates current, whilst the solid line shows the differential of the current taken with respect to $V_{b}$; of interest because it can be compared directly with experimental results.  The panels on the right display the relative alignment of the monolayer and bilayer graphene bands at bias voltages corresponding to distinctive features in the current.  Feature (a) arises due to onset of resonant tunnelling from the high energy bilayer valence band to monolayer valence band as the monolayer chemical potential crosses their point of intersection.  This can be identified in the corresponding inset, where solid lines denote occupied electron states and dashed lines imply the unoccupied ones above the chemical potential of the relevant layer.  Feature (b) occurs as the bilayer chemical potential reaches the low energy bilayer conduction band minimum, as indicated by a solid black horizontal line for clarity.  This reduces the number of tunnelling states available and leads to a decrease in the magnitude of the tunnelling current, hence producing a peak in $dI/dV_{b}$.  The group of positive and negative peaks around feature (c) are generated when, first, the bilayer chemical reaches the low energy bilayer valence band maxima resulting in a loss of tunnelling states.  The curve then exhibits a strong negative peak near to where the monolayer Dirac point crosses the low energy bilayer valence band, resulting in a large range of wavevectors which can contribute to tunnelling due to inelastic broadening, $\alpha$.  Once this condition is no longer satisfied, the differential conductance experiences another peak as the magnitude of the negative tunnelling current reduces rapidly.  Features (d) and (d$^{'}$) bound a region which exhibits zero tunnelling current.  Each peak manifests the onset of resonant tunnelling as the monolayer chemical potential passes through the point of intersection between the monolayer and low energy bilayer valence and conduction bands respectively. Feature (e) is produced when the bilayer chemical potential drops below the energy value associated with the intersection between the monolayer and low energy bilayer valence bands.  Finally, the peak in differential conductance marked (f) is produced as the monolayer chemical potential rises above the intersection between the monolayer and high energy bilayer conduction bands.  This results in an onset of resonant tunnelling into this available energy band.	
	
	\begin{figure}[htbp]
		\includegraphics[width=0.48\textwidth]{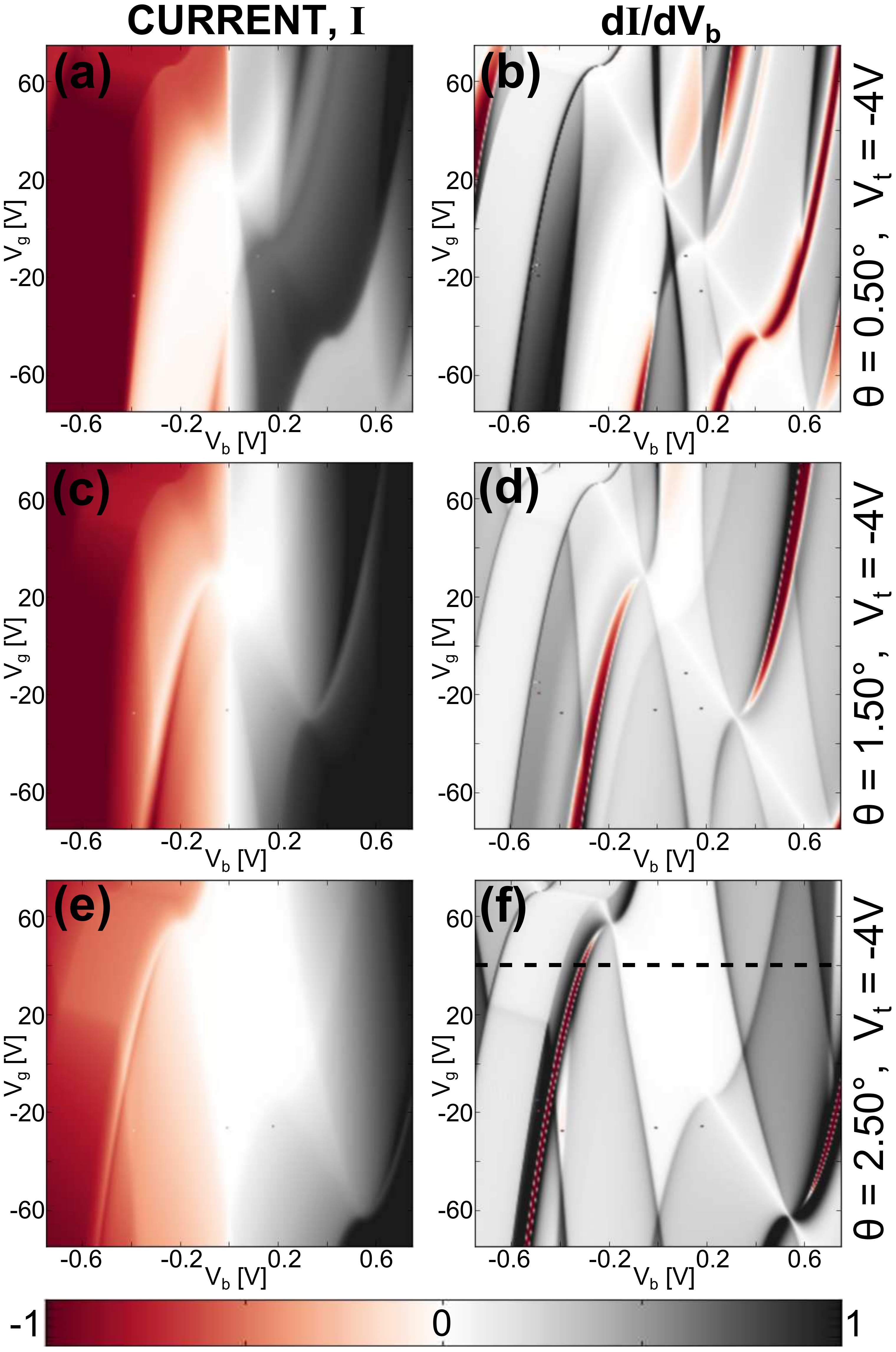}
		\caption{
			Colour maps of current density (left) and differential conductance (right) normalised to the highest data value for a series of misalignment angles.  The dashed line identifies the location of the cut-through illustrated in Fig.\ref{current_cut}.\label{currentResults_2D}}
	\end{figure}
	
	The position of each of these features varies with both $V_{g}$ and $\theta$ as is evident from Fig.\ref{currentResults_2D}, which shows how the characteristics of the current and its differential move as we alter the misalignment angle between the graphene electrodes.  Studying these images, we note that varying the misalignment angle effectively acts to restrict the number of characteristics available within our energy/voltage range.  We can also see that the distinctive negative peak corresponding to feature (c) in Fig.\ref{current_cut} persists throughout a large gate voltage range, producing a narrow band of NDC accessible even at extremely low bias voltages.  
	
	The central region of Fig.\ref{currentResults_2D}(f) (and also appearing in a smaller voltage range for the other angles) represents an area in the I/V characteristics that exhibits zero conductance.  This is bounded by four characteristic curves of the type indicated in Fig.\ref{current_cut}(d).  The lower/upper curves satisfy the condition that the bilayer chemical potential equals the energy value of the intersection between the monolayer and low energy conduction/valence bands, whilst the curves to the left/right occur at the voltage values required for the monolayer chemical potential to reach the intersection between the monolayer and low energy bilayer valence/conduction bands.
	
	To conclude, we have developed a method for describing tunnelling between bilayer and monolayer graphene electrodes separated by an insulating layer of hBN with both bottom and top gates to achieve a finer degree of control.  Within this, we have included a self-consistent description of the system's electrostatics which allows us to control the device characteristics by tuning three independent voltages across it.  We have shown that a narrow band of negative differential conductance is accessible at relatively low voltages, which opens possibilities to use monolayer-bilayer graphene tunnelling devices for non-linear high-frequency generators, as in Ref.\cite{mishchenko_nature_2014, gaskell_arxiv_2015}.\\
	
	We would like to thank K.~Novoselov, A.~Mishchenko and D.~Ghazaryan for usefull discussions. We acknowledge financial support from the EU Graphene Flagship Programme and CDT NOWNANO.


\begin{thebibliography}{99}
		\bibitem{ponomarenko_natPhys_2011}L.A. Ponomarenko, A.K. Geim, A.A. Zhukov, R. Jalil, S.V. Morozov, K.S. Novoselov, I.V. Grigorieva, E.H. Hill, V.V. Cheianov, V.I. Fal’Ko, K. Watanabe, T. Taniguchi, \& R.V. Gorbachev, Nature Physics 7, 958 (2011).
		\bibitem{lee_appPhysLett_2011}G.-H. Lee, Y.-J. Yu, C. Lee, C. Dean, K.L. Shepard, P. Kim, and J. Hone, Applied Physics Letters 99, 243114. (2011)
		\bibitem{britnell_nanoLett_2012}L. Britnell, R.V. Gorbachev, R. Jalil, B.D. Belle, F. Schedin, M.I. Katsnelson, L. Eaves, S.V. Morozov, A.S. Mayorov, N.M.R. Peres, A.H.C. Neto, J. Leist, A.K. Geim, L.A. Ponomarenko, \& K.S. Novoselov, Nano Letters 12, 1707 (2012).
		\bibitem{yang_science_2012}H. Yang, J. Heo, S. Park, H.J. Song, D.H. Seo, K.-E. Byun, P. Kim, I. Yoo, H.-J. Chung, \& K. Kim, Science 336, 1140 (2012).
		\bibitem{haigh_natMat_2012}S.J. Haigh, A. Gholinia, R. Jalil, S. Romani, L. Britnell, D.C. Elias, K.S. Novoselov, L.A. Ponomarenko, A.K. Geim, \& R. Gorbachev, Nature Materials 11, 764 (2012).
		\bibitem{gorbachev_natPhys_2012}R.V. Gorbachev, A.K. Geim, M.I. Katsnelson, K.S. Novoselov, T. Tudorovskiy, I.V. Grigorieva, A.H. Macdonald, S.V. Morozov, K. Watanabe, T. Taniguchi, and L.A. Ponomarenko, Nature Physics 8, 896 (2012).
		\bibitem{britnell_science_2013}L. Britnell, R.M. Ribeiro, A. Eckmann, R. Jalil, B.D. Belle, A. Mishchenko, Y.- J. Kim, R.V. Gorbachev, T. Georgiou, S.V. Morozov, A.N. Grigorenko, A.K. Geim, C. Casiraghi, A.H.C. Neto, \& K.S. Novoselov, Science 340, 1311 (2013).
		
		\bibitem{britnell_science_2012}L. Britnell, R.V. Gorbachev, R. Jalil, B.D. Belle, F. Schedin, A. Mishchenko, T. Georgiou, M.I. Katsnelson, L. Eaves, S.V. Morozov, N.M.R. Peres, J. Leist, A.K. Geim, K.S. Novoselov, \& L.A. Ponomarenko, Science 335, 947 (2012).
		\bibitem{georgiou_natNano_2013}T. Georgiou, R. Jalil, B.D. Belle, L. Britnell, R.V. Gorbachev, S.V. Morozov, Y.-J. Kim, A. Gholinia, S.J. Haigh, O. Makarovsky, L. Eaves, L.A. Ponomarenko, A.K. Geim, K.S. Novoselov, \& A. Mishchenko, Nature Nanotechnology 8, 100 (2013).
		\bibitem{ponomarenko_appPhys_2013}L.A. Ponomarenko, B.D. Belle, R. Jalil, L. Britnell, R.V. Gorbachev, A.K. Geim, K.S. Novoselov, A.H.C. Neto, L. Eaves, \& M.I. Katsnelson, Journal Of Applied Physics 113, 136502 (2013).
		\bibitem{britnell_natComm_2013}L. Britnell, R.V. Gorbachev, A.K. Geim, L.A. Ponomarenko, A. Mishchenko, M.T. Greenaway, T.M. Fromhold, K.S. Novoselov, \& L. Eaves, Nature Communications 4, 1794 (2013).
		\bibitem{mishchenko_nature_2014}Mishchenko, A., J. S. Tu, Y. Cao, R. V. Gorbachev, J. R. Wallbank, M. T. Greenaway, V. E. Morozov, S. V. Morozov, M. J. Zhu, S. L. Wong, F. Withers, C. R. Woods, Y-J. Kim, K. Watanabe, T. Taniguchi, E. E. Vdovin, O. Makarovsky, T. M. Fromhold, V. I. Fal'ko, A. K. Geim, L. Eaves, \& K. S. Novoselov.
		Nature Nanotech Nature Nanotechnology 9, 808 (2014).
		\bibitem{fallahazad_nanoLet_2009}B. Fallahazad, K. Lee, S. Kang, J. Xue, S. Larentis, C. Corbet, K. Kim, H.C.P. Movva, T. Taniguchi, K. Watanabe, L.F. Register, S.K. Banerjee, \& E. Tutuc, Nano Letters 15, 428 (2015).
		
		\bibitem{geim_natMat_2007}A.K. Geim \& K.S. Novoselov, Nature Materials 6, 183 (2007).
		\bibitem{neto_revModPhys_2009}A.H.C. Neto, F. Guinea, N.M.R. Peres, K.S. Novoselov, \& A.K. Geim, Reviews Of Modern Physics 81, 109 (2009).
		\bibitem{mccann_repProgPhys_2013}E. Mccann \& M. Koshino, Reports On Progress in Physics 76, 056503 (2013).
		\bibitem{yang_nano_2010}X. Yang, G. Liu, A.A. Balandin, \& K. Mohanram, ACS Nano 4, 5532 (2010).
		\bibitem{wang_nanoLett_2015}X. Wang, X. Jiang, T. Wang, J. Shi, M. Liu, Q. Zeng, Z. Cheng, \& X. Qiu, Nano Letters 15. 3212 (2015).
		\bibitem{gaskell_arxiv_2015}J. Gaskell, L. Eaves, K.S. Novoselov, A. Mishchenko, A.K. Geim, T.M. Fromhold, \& M.T. Greenaway. arXiv:1506.05053v1 (2015)
		\bibitem{zhang_nature_2009}Y. Zhang, T.-T. Tang, C. Girit, Z. Hao, M.C. Martin, A. Zettl, M.F. Crommie, Y.R. Shen, \& F. Wang, Nature 459, 820 (2009).
		\bibitem{ohta_scienceMag_2006}T. Ohta, A. Bostwick, T. Seyller, K. Horn, and E. Rotenberg, Science 313, 951 (2006).
		\bibitem{dean_natNano_2010}C.R. Dean, A.F. Young, I. Meric, C. Lee, L. Wang, S. Sorgenfrei, K. Watanabe, T. Taniguchi, P. Kim, K.L. Shepard, and J. Hone, Nature Nanotechnology 5, 722 (2010).
		\bibitem{bistritzer_physRevB_2010}R. Bistritzer, A.H. MacDonald, Physical Review B 81, 245412 (2010).
		\bibitem{mele_physRevB_2011}E.J. Mele, Physical Review B 84, 235439 (2011).
		\bibitem{santos_physRevB_2012}J.M.B Lopes dos Santos, N.M.R Peres, A.H. Castro Neto, Physical Review 86, 155449 (2012).
		\bibitem{kindermann_physRevB_2012} M. Kindermann, B. Uchoa, D.L. Miller, Physical Review B 86, 115415 (2012).
		\bibitem{wallbank_thesis} J.~R.~Wallbank, \emph{Electronic Properties of Graphene Heterostructures with Hexagonal Crystals} (Springer PhD Thesis Series, Springer, 2014).
		\bibitem{moon_physRevB_2014}P. Moon, M. Koshino, Physical Review B 90, 155406 (2014).
		
		\bibitem{novoselov_natPhys_2006}K.S. Novoselov, E. Mccann, S.V. Morozov, V.I. Fal’Ko, M.I. Katsnelson, U. Zeitler, D. Jiang, F. Schedin, and A.K. Geim, Nature Physics 2, 177 (2006).
		\bibitem{mccann_physRevB_2006}McCann, E. Physical Review B 74, 161403 (2006).
				 
		\bibitem{footnote} An electron scattered by hBN reciprocal lattice vector, $\vec{g}_{BN}$, while tunnelling between BZ corners $\vec K^{\text{BL}}_{n,\xi}$ and $\vec K^{\text{ML}}_{m,\xi}$ on the BLG/MLG (labelled according to Fig.~\ref{theStackWithBZ}), has its valley momentum shifted by $\delta \vec {p'}_{n,m,\xi}= \vec K^{\text{ BL}}_{n,\xi} - \vec g_{BN} - \vec K^{\text{ ML}}_{m,\xi}$ \cite{wallbank_thesis}. 		
		The initial/final state energies in this process are $\epsilon\sim v|\delta \vec {p'}_{n,m,\xi}|$ and will therefore be Pauli blocked, for all achievable levels of electrostatic doping, unless $|\delta \vec {p'}_{n,m,\xi}|\ll |\vec K_+|$. 
		This condition can only be satisfied with $\vec{g}_{BN}\neq0$ if the hBN layer is crystallographically well aligned with the two graphene layers, and the resulting addtional tunnelling process can be included into Eq.~\eqref{fermi_golden_rule}, by replacing $\delta \vec{p'}_{n,\xi}\rightarrow\delta \vec {p'}_{n,m,\xi}$ throughout, $n\rightarrow m$ in $G_{ML}$, and $\sum_n\rightarrow\sum_{n,m}$.
		 
		
	\end{thebibliography}
\end{document}